\documentclass[3p,times]{elsarticle}

\usepackage{ecrc}
\usepackage{sidecap}

\usepackage{epstopdf}

\volume{00}

\firstpage{1}

\journalname{Nuclear Physics A}

\runauth{Author1 et al.}

\jid{nupha}

\jnltitlelogo{Nuclear Physics A}

\usepackage{graphicx}
\usepackage{amsmath,amssymb}

\usepackage{multicol}

\begin{document}

\begin{frontmatter}

\title{An Overview of STAR Experimental Results}

\author{Nu Xu$^{1,2}$ (for the STAR Collaboration)}
\address{$^1$Nuclear Science Division, Lawrence Berkeley National Laboratory, Berkeley, CA 94720, USA }
\address{$^2$Key Laboratory of Quarks and Lepton Physics (MOE) and Institute of Particle Physics, Central China Normal University, Wuhan, 430079, China}


\begin{abstract}
With large acceptance and excellent particle identification, STAR is one of the best mid-rapidity collider experiments for studying high-energy nuclear collisions. The STAR experiment provides full information on initial conditions, properties of the hot and dense medium as well as the properties at freeze-out.
In Au+Au collisions at $\sqrt{s_{NN}} = 200$ GeV, STAR's focus is on the nature of the sQGP produced at RHIC. In order to explore the properties of the QCD phase diagram, since 2010, the experiment has collected sizable data sets of Au+Au collisions at the lower collision energy region where the net-baryon density is large. 

At the 2014 Quark Matter Conference, the STAR experiment made 16 presentations that cover physics topics including {\it collective dynamics}, {\it electromagnetic probes}, {\it heavy flavor},  {\it initial state physics}, {\it jets},  {\it QCD phase diagram}, {\it thermodynamics and hadron chemistry}, and {\it future experimental facilities, upgrades, and instrumentation} [1-16]. In this overview we will highlight a few results from the STAR experiment, especially those from the recent measurements of the RHIC beam energy scan program. At the end, instead of a summary, we will discuss STAR's near future physics programs at RHIC. 
\end{abstract}

\end{frontmatter}


\section{Introduction}
\label{intro}

Initially when Lee and Wick first proposed studying the high-energy nuclear collisions their goal was to create a new form of nuclear matter called the Quark-Gluon Plasma (QGP)~\cite{tdlee}. It turns out that the net-baryon density as well as the temperature strongly depend on the colliding energy, therefore high-energy collisions are also very effective for studying  the QCD phase diagram. At mid-rapidity, the higher the collision energy the lower the net-baryon density~\cite{pbm}. In ultra-relativistic heavy-ion collisions, where the net-baryon density is close to zero, the strongly coupled QGP has been observed~\cite{rhicwp,aliceqm14} at both RHIC and the LHC. The properties of the medium created in such collisions show a strong opacity to colored objects and small ratio of shear viscosity over entropy density\cite{rhicwp1}. 

While the study of the nature of the sQGP continues in the high energy region, the first RHIC beam energy scan (RHIC BESI) program, all with Au+Au collisions, was started in 2010.  The main motivation there is to systematically explore\footnote{The STAR experiment has a large uniform acceptance at mid-rapidity covering $|\eta| \le 1.1$ and excellent particle identification. Its performance does not change as a function of collision energy. It is an ideal detector system for studying the thermodynamic properties of the medium created in collisions especially involving the fluctuation/correlation analysis. } the nuclear matter phase structure, the emergent property with the QCD degrees of freedom, at higher baryon region. As of May 2014, when the RHIC BESI was concluded, we have covered the beam energies of $\sqrt{s_{NN}}= 39, 27, 19.6, 14.5, 11.5$, and 7.7 GeV corresponding to a range of chemical potentials\footnote{At a given energy, especially at the lower beam energy, the value of the chemical potential varies with collision centrality. These values were extracted from the central (0-5\%) Au+Au collisions.}  of 110 $\le \mu_B \le$ 420 MeV. Including the previously measured energies of $\sqrt{s_{NN}}= 200, 130$ and 62.4 GeV, RHIC has provided nine energies for heavy-ion beams. Although most of the measured quantities change smoothly within the above energy range, one noticed that in Au+Au collisions above $\sqrt{s_{NN}}= 39$ GeV, the thermodynamic properties of the medium are very similar to those of the collisions at the top energy 200 GeV, implying that the partonic interactions dominate the dynamics. In the lower energy collisions, the deviation from the sQGP becomes clearer indicating hadronic degrees of freedom play a more important role.  In this review we will highlight some of the observations and discuss the required future measurements needed to gain further insight into the QCD phase diagram.


\section{Penetrating Probes: Study the QGP Medium Properties}

\subsection{Results on di-electron production}
\label{sub:die}

The di-leptons have long been considered as a penetrating probe for studying the collision dynamics in high-energy nuclear collisions and they also make a bulk probe as leptons are continuously emitted during the evolution. This bulk-penetrating probe will potentially provide information of QGP direct radiation in the intermediate mass region $1\le M_{ll} \le 3$ GeV/c$^2$ and chiral dynamics in the low mass region $M_{ll} \le 1$ GeV/c$^2$. 

\begin{SCfigure}
\includegraphics*[width=7.cm]{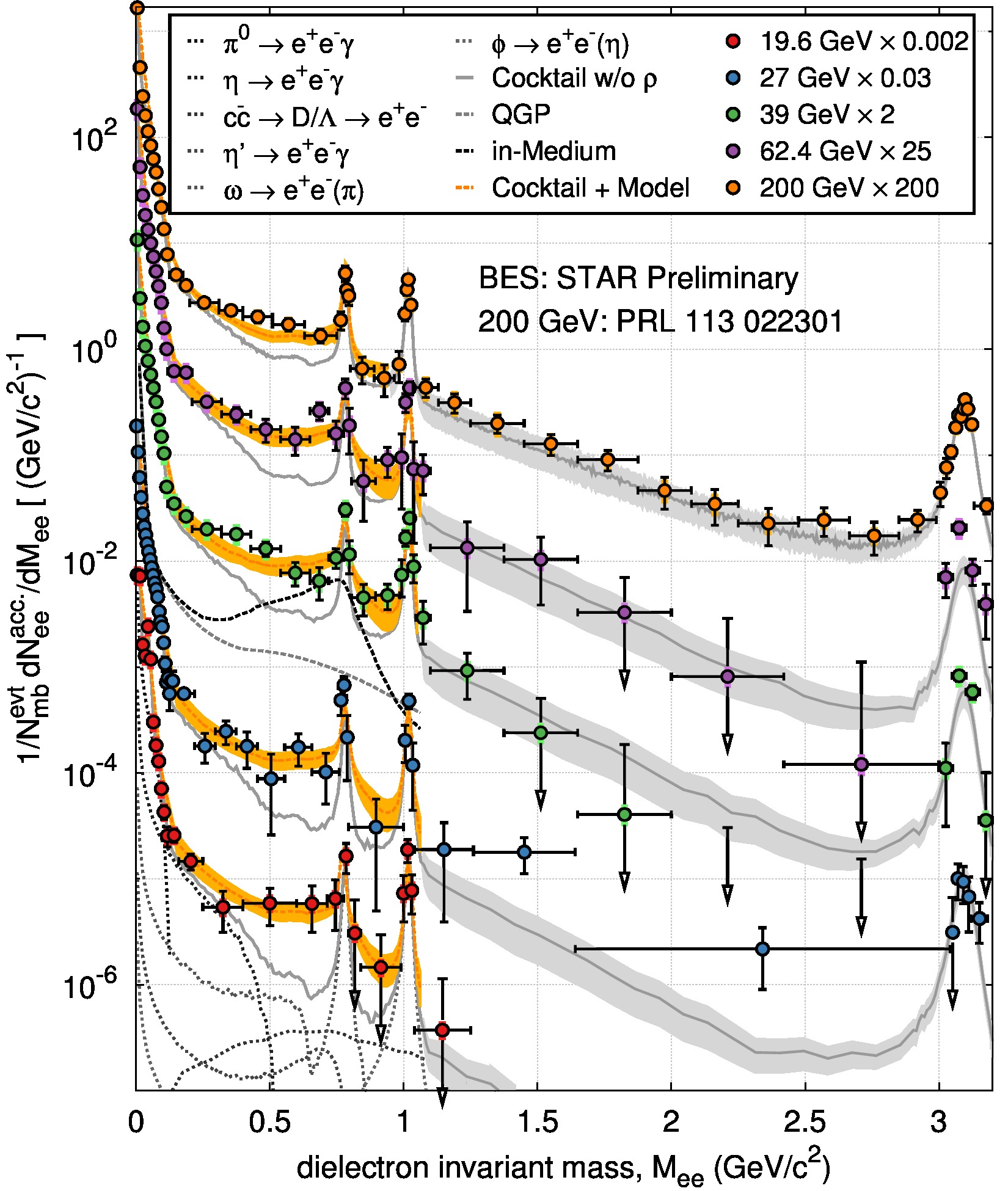}
\caption{Collision energy dependence of di-electron invariant mass distributions~\cite{phuck} from minimum bias Au+Au collisions at
  200, 62.4, 39, 27 and 19.6 GeV, from top to bottom, respectively. In the low mass region, the yellow-bands are the results from hadronic cocktails plus in-medium model calculations. The gray-bands are the estimated systematic error bars.}
\label{fig:die}
\end{SCfigure}

With its large, uniform acceptance and excellent particle identification capability, the STAR experiment is in a unique position to make a systematic study of di-electron production in the RHIC BESI program. Figure~\ref{fig:die} shows the efficiency-corrected di-electron invariant mass spectra from minimum-bias Au+Au collisions at $\sqrt{s_{NN}}=200, 62.4, 39, 27$, and 19.6 GeV~\cite{phuck}. The 200 GeV result has been published in Ref.~\cite{die2014}. The yellow-bands in the low mass region represent the results of hadronic cocktails plus in-medium model calculations~\cite{ralf00}. The driving ingredient in the model calculation is the meson-baryon re-scatterings during the evolution. Within the error bars, the model results are consistent with the measured data. The collisional broadening of the $\rho$-mesons leads to the enhancement in this mass region.  At lower beam energies, the baryon over meson ratio will increase due to strong stopping, therefore one would expect further increase in the observed enhancement~\cite{phuck,besII}.

\subsection{Results on heavy flavor production}
\label{sub:hf}

\begin{SCfigure}
\includegraphics*[width=7.5cm]{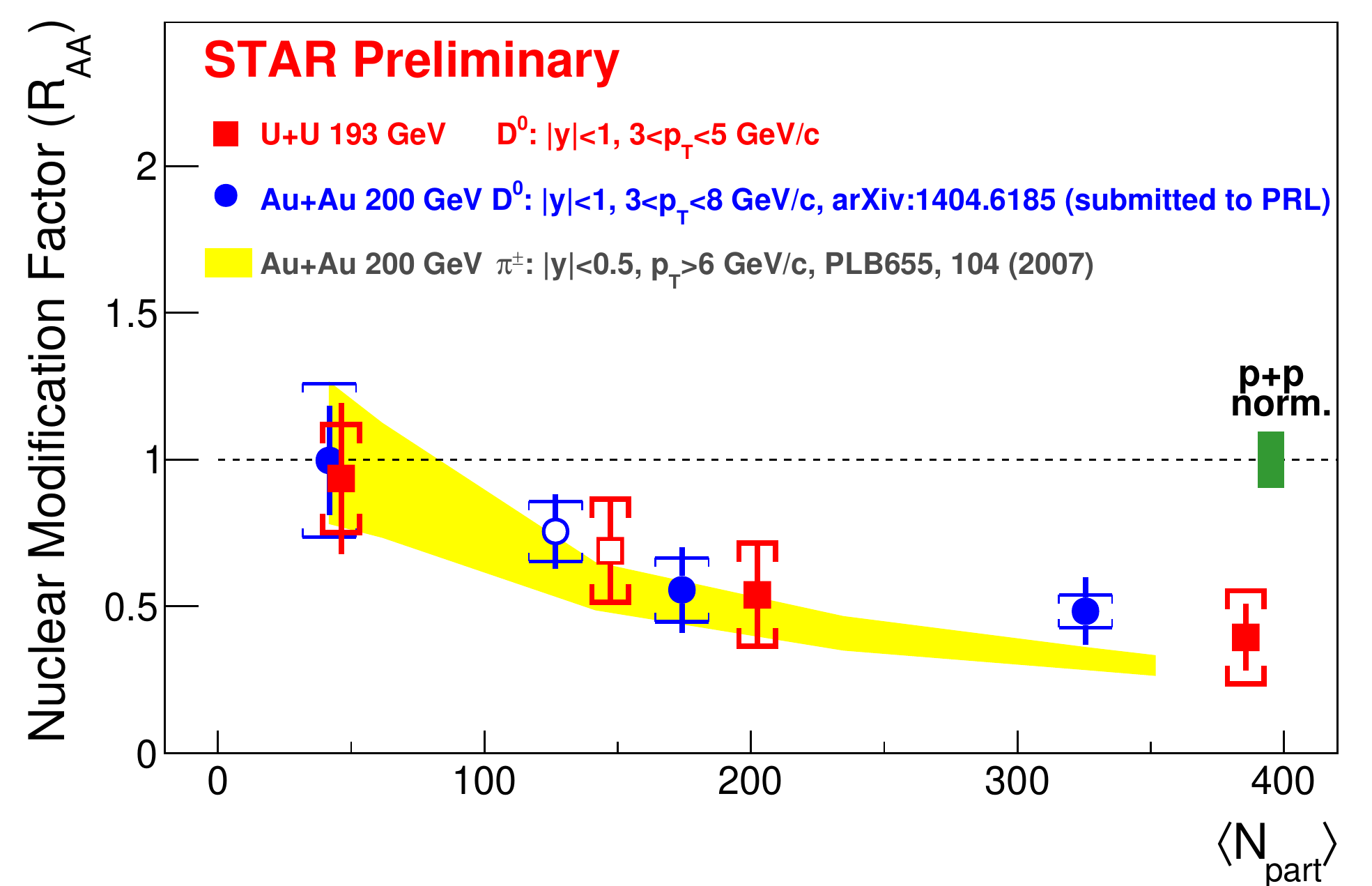}
\caption{Collision centrality dependence of the nuclear modification factor $R_{AA}$ for $D^0$ mesons~\cite{zyye}.  Blue and red symbols are for 
 results from $\sqrt{s_{NN}}=200$ GeV Au+Au and $\sqrt{s_{NN}}=193$ GeV U+U collisions, respectively. The vertical error bars are statistical errors while the caps are the estimated systematic errors. The yellow band is the results of large $p_T$ pion $R_{AA}$ from $\sqrt{s_{NN}}=200$ GeV  Au+Au collisions~\cite{starpi07}. The normalization uncertainty from the p+p collisions is shown as the vertical green bar.}
\label{fig:hf}
\end{SCfigure}

Heavy flavor here refers to hadrons with at least one heavy quark ($c$ or $b$). Due to the large masses of heavy quarks, thermal heavy flavor production is suppressed leaving  only initial hard scattering to create heavy flavors in high-energy nuclear collisions at RHIC. The heavy flavors are therefore a useful tool for studying the medium properties.  In Fig.~\ref{fig:hf}, the nuclear modification factors $R_{AA}$ of $D^0$ mesons are shown as a function of collision centrality,  represented by the average number of participant nucleons $\langle N_{part} \rangle$~\cite{zyye}. The blue and red symbols represent the results from $\sqrt{s_{NN}}=200$ GeV Au+Au~\cite{stard0} and $\sqrt{s_{NN}}=193$ GeV U+U collisions, respectively. For comparison, the large transverse momentum pion result is also shown as the yellow band~\cite{starpi07}.  

As illustrated in Fig.~\ref{fig:hf}, charm hadrons suffer similar amount of energy loss compared to that of the light flavor hadrons implying that the hot/dense medium, created in high-energy nuclear collisions at RHIC, is opaque to light- as well as heavy flavors. This conclusion is also consistent with non-photonic electron $R_{AA}$ at high $p_T$~\cite{stareraa}. The increase in density in the heavier colliding system U+U is of the order of 20\%~\cite{hwang,hiroshi09}. The trend of the suppression from U+U collisions in the most central bin is consistent with the expectation.

\begin{SCfigure}
\includegraphics*[width=6.5cm]{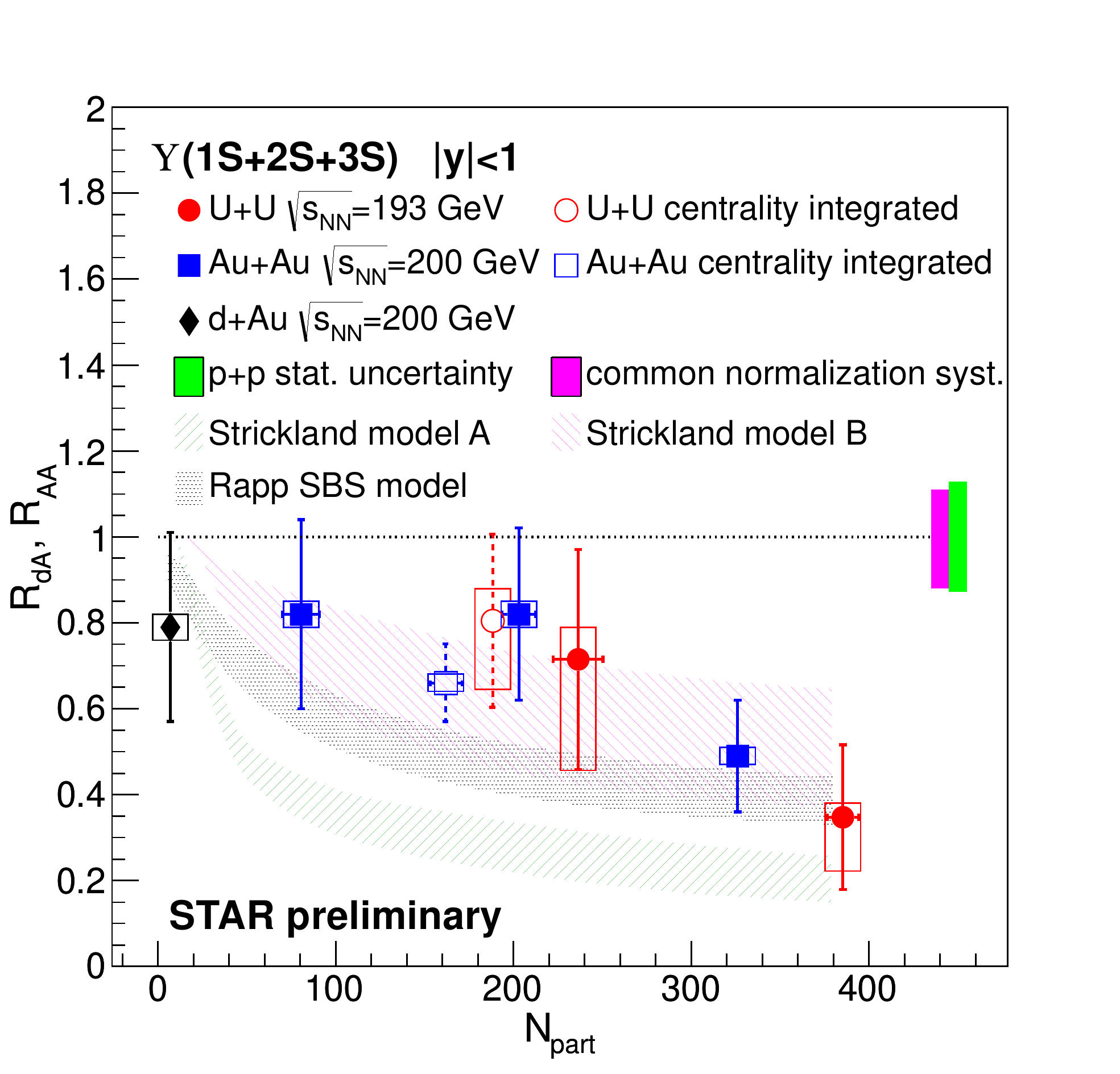}
\caption{Collision centrality dependence of the nuclear modification factor $R_{AA}$ for $\Upsilon$(1S+2S+3S) in $\sqrt{s_{NN}}=200$ GeV d+Au (diamonds) and Au+Au (squares) and $\sqrt{s_{NN}}=193$ GeV U+U (circles)  collisions~\cite{wmzha}. Statistical error of $\Upsilon$s from the p+p collisions and the overall normalization systematic error are shown as green and cyan bars, respectively.  Hatched bands are model calculations as discussed in the text.}
\label{fig:upsilon}
\end{SCfigure}

Nuclear modification factors of mid-rapidity $\Upsilon$(1S+2S+3S) in d+Au, Au+Au and U+U collisions are presented in Fig.~\ref{fig:upsilon} with respect to the number of participants. Similar to the open-charm hadron case, the trend observed in Au+Au is generally continued in the U+U data, with an $R_{\rm{AA}}=0.35\pm 0.17 (stat.) ^{+0.03}_{-0.13} (syst.)$ measured for the 10\% most central U+U collisions. The STAR experiment has published the results for $\sqrt{s_{NN}}=200$ GeV p+p, d+Au and Au+Au collisions~\cite{Adamczyk:2013poh}. The observed suppression for the  $\Upsilon$(1S) state cannot be explained by the cold nuclear matter effect alone. The final-state interactions lead to further suppression. The observed yields of the excited states are consistent with a complete suppression scenario.  The model of Strickland and Bazow~\cite{Strickland:2011aa} which incorporates lattice QCD results on screening and broadening of bottomonium, is the pink-hatched band in the figure. This scenario, with a potential based on heavy quark internal energy, is consistent with the observations. While the free energy based scenario, green-hatched band, shows over-suppression so it is disfavoured. The strong binding scenario, black-hatched band in the figure, proposed by Emerick, Zhao, and Rapp~\cite{Emerick:2011xu}, which includes possible CNM effects in addition, is also consistent with STAR results.

\section{Results from the First Beam Energy Scan (RHIC BESI): Exploring the QCD Phase Structure}
\subsection{Results on chemical freeze-out}

\begin{SCfigure}
\includegraphics*[width=10cm]{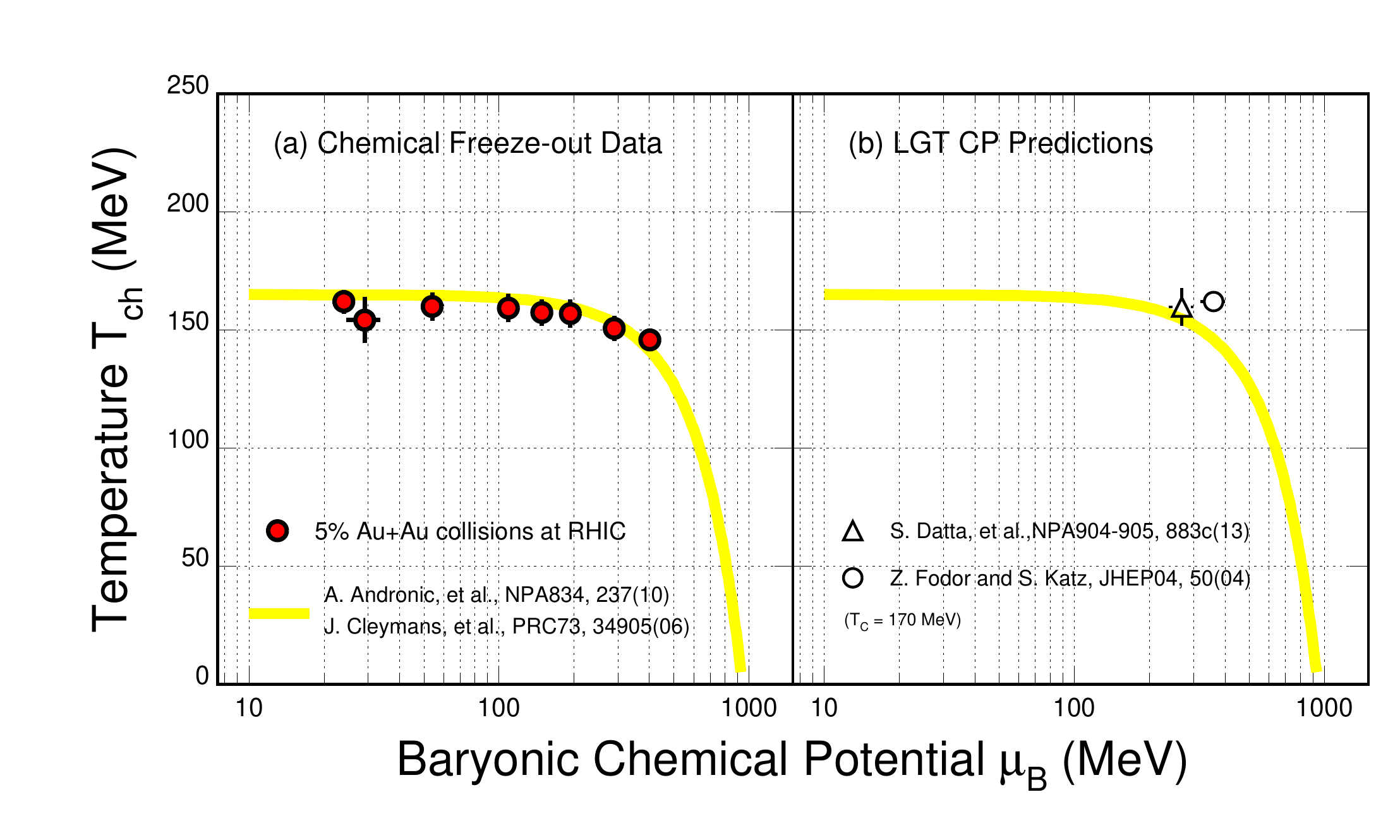}
\caption{(a) Chemical freeze-out temperature  ($T_{\mathrm {ch}}$) versus baryonic chemical potential ($\mu_{\mathrm {B}}$) obtained from a statistical 
  model~\cite{aandronic10}  fit to yields of hadrons produced in 0-5\% central Au+Au collisions at RHIC. The yellow bands
  are the empirical results from fitting experimental data acquired prior to the RHIC BESI program by a statistical model. (b) The positions of the QCD critical point from two different lattice gauge theory calculations in the  $T_{\mathrm {ch}}$ versus $\mu_{\mathrm {B}}$ plane are shown.}
\label{fig:cfo}
\end{SCfigure}

Assuming that the medium, created in high-energy nuclear collisions, has reached a thermal equilibrium and thus can be expressed in terms of a few thermodynamic parameters, the location of the hot and dense system at freeze-out is then well defined in the phase diagram.  Figure~\ref{fig:cfo}(a) shows the results of the chemical freeze-out\footnote{We often discuss two freeze-out situations in high-energy nuclear collisions. (1) Chemical freeze-out: At the moment when the inelastic scatterings cease. The population of all hadron states are fixed and no further changes in hadron yields after the freeze-out.  (2) Kinetic freeze-out: At the moment when the elastic scatterings stop. After that the freeze-out particle momentum distributions no longer change. Some of the discussions can be found in \cite{lkumar}.} parameters from the 0-5\% central Au+Au collisions. A systematic analysis of the centrality dependence in the RHIC BESI at STAR has also been carried out and the results can be found in Ref.~\cite{lkumar13}. The yellow line in the plot is an empirical parameterization of the freeze-out conditions~\cite{aandronic10,jcleymans06} in central heavy-ion collisions. As shown in the figure, the STAR results cover a wide range of the baryonic chemical potential $20 \le \mu_B \le 420$ MeV. 

Theoretical estimations for the QCD critical point~\cite{zfodor04,sdatta13} are shown in Fig.~\ref{fig:cfo}(b) as open symbols. A value of $T_c = 170$ MeV was used in the theoretical estimate. Although the dynamic backgrounds between data and model calculations may be different and a direct comparison may be misleading, the region indicated by the model calculation is indeed covered by the RHIC BES program. Clearly, the high net-baryon region is favored for the search for the QCD critical point. 

\subsection{Results on flow measurements}

The ground state of the uranium nucleus is deformed with a prolate shape of $\beta_2 \sim 0.28$. Compared to that of Au+Au collisions, the central U+U collisions not only give rise to a higher energy density~\cite{hiroshi09},  they also allow us to separate two extreme collision configurations: tip-tip versus body-body. Information on the initial conditions may be extracted by analyzing the extreme orientations~\cite{svoloshin10}. 

\begin{SCfigure}
\includegraphics*[width=10cm]{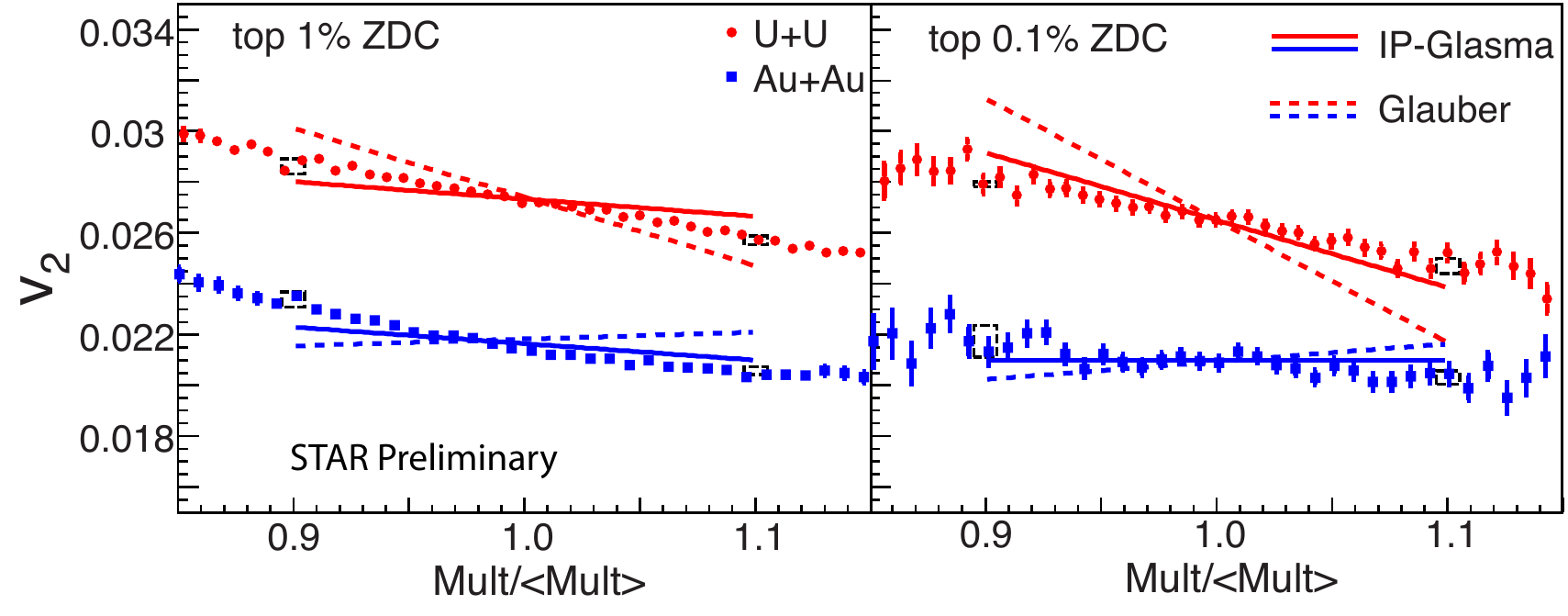}
\caption{The $v_2$ of charged particles as a function of the normalized multiplicity~\cite{hwang} for $\sqrt{s_{NN}}=200$ GeV  Au+Au (blue symbols) and $\sqrt{s_{NN}}=193$ GeV U+U (red symbols) collisions. The left and right panels show the results of top 1\% and 0.1\% central events measured by the Zero Degree Calorimeter measured, respectively. The lines represent Glauber (dashed lines) and IP-Glasma (solid lines) simulations.
The small boxes indicate systematic uncertainties due to the efficiency corrections for the multiplicity measurements.}
\label{fig:flow}
\end{SCfigure}

Figure~\ref{fig:flow} shows the $v_2$ of charged particles as a function of the normalized multiplicity~\cite{hwang} for $\sqrt{s_{NN}}=200$ GeV  Au+Au (blue symbols) and $\sqrt{s_{NN}}=193$ GeV U+U (red symbols) collisions. The left and right panels show the results of top 1\% and 0.1\% central events selected by the  Zero Degree Calorimeter, respectively. As one can see in the right plot, the slope for the top 0.1\% ZDC U+U collision is steeper compared to that of the 1\% case indicating that in such central collisions, the geometrical effect is dominant rather than the fluctuations, as for Au+Au collisions~\cite{hwang}. Model calculations are compared with the $v_2$ measurements. It is clear that gluon-saturation-based calculations, IP-Glasma~\cite{bschenke14} shown as solid lines in Fig.~\ref{fig:flow}, fit better to the measurements.

\subsection{Results on high moment analysis }

As mentioned above, a smooth cross-over is expected at the vanishing baryonic chemical potential. At the higher net-baryon region, however, model calculations have speculated the existence of a first-order phase transition. In that case thermodynamic principles suggest that there should be a critical point in QCD matter where the first-order phase transition ends and the transition becomes a cross-over~\cite{aoki06},  at which point the phase boundaries effectively cease to exist.
The characteristic experimental signature of the QCD critical point  is large fluctuations in event-by-event multiplicity distributions of conserved quantities such as net-charge, net-baryon number, and net-strangeness. The variances of these distributions, $\langle (\delta N)^{2} \rangle$, are proportional to the square of the correlation length $\xi$.

Model studies have shown that higher order  moments ($\langle (\delta N)^{3} \rangle$ $\sim$ $\xi^{4.5}$ and $\langle (\delta N)^{4} \rangle$ $\sim$ $\xi^{7}$) have stronger dependences on $\xi$ than the variance and thus have higher sensitivity~\cite{mstephanov0911, masakawa09}.
Another salient feature of the moments is that they  can be reconstructed from susceptibilities ($\chi$)~\cite{mcheng09}. For example,  $\kappa$$\sigma^2$ = $\frac{\chi^{(4)}}{\chi^{(2)}/T^2}$~\cite{rgavai11}. Hence  a comparison of the experimental data can be made directly to QCD model calculations~\cite{rgavai11,sgupta11}. In high-energy nuclear collisions, the created system has a finite size, time span and the number of particles are also finite. Instead of a point in the phase diagram, one may observe a critical region in which large fluctuation might be observed. On the other hand, in the absence of a critical point, the hadron resonance gas model~\cite{Garg:2013ata} suggests that the $\kappa$$\sigma^2$ values will be close to unity and have a monotonic dependence on $\sqrt{s_{\mathrm {NN}}}$ ~\cite{Luo:2013bmi,xfluo2014}.

\begin{SCfigure}
\includegraphics*[width=6.5cm]{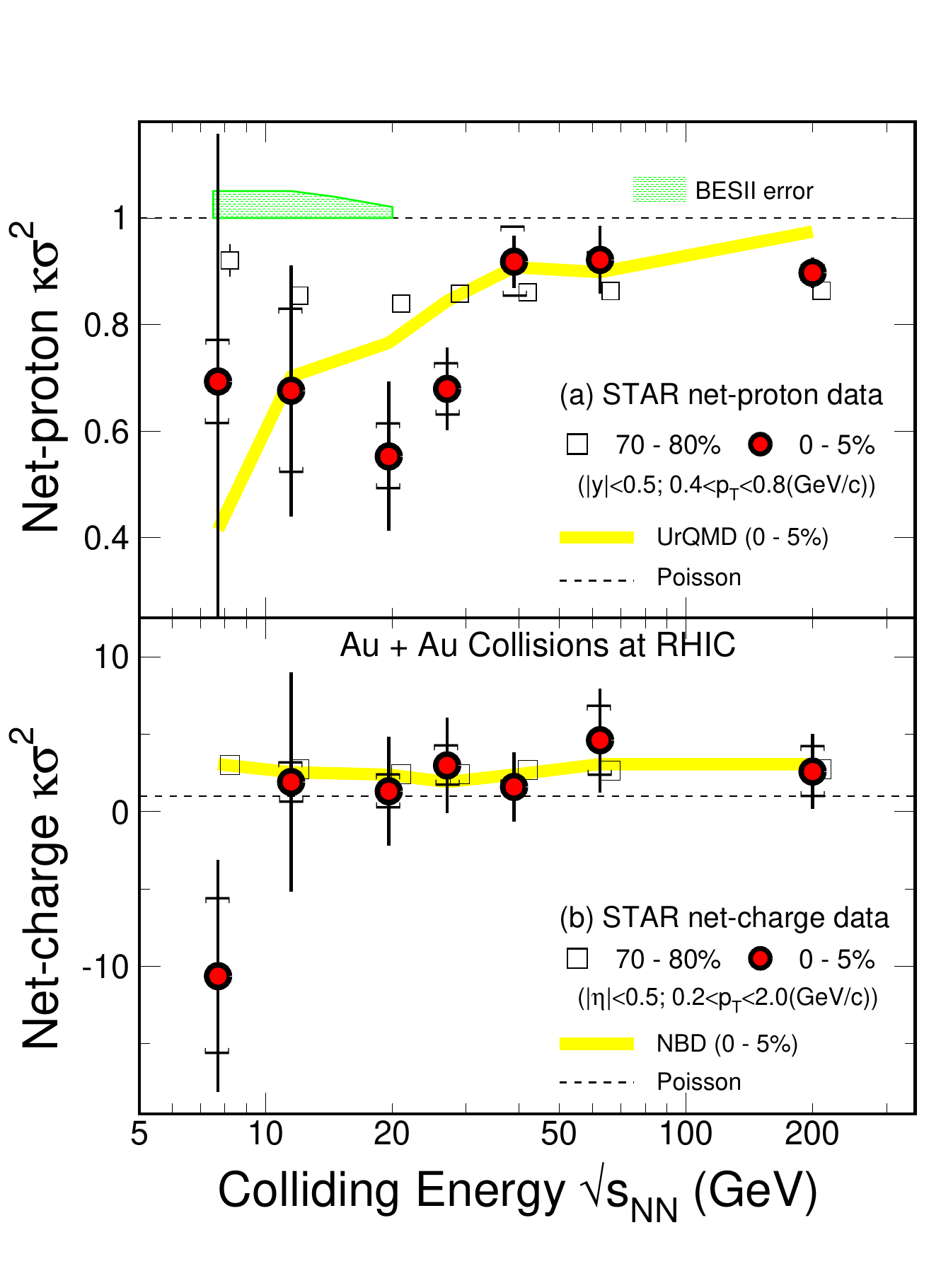}
\caption{Collision energy dependence of net-proton (top panel)~\cite{netp2014}  and net-charge (bottom
panel)~\cite{netq2014} high moments $\kappa \sigma^2$ from Au+Au collisions at RHIC. The red solid circles 
correspond to 0-5\% central collisions and the open squares represent 70-80\% peripheral collisions. The vertical 
error bars are statistical and the caps correspond to systematic errors. The yellow solid line in the top panel
represents 0-5\% central Au+Au collision results from UrQMD simulations and the yellow solid line in the bottom 
panel is the result of negative binomial statistics. The dashed lines in both panel represent the Poisson statistics.
The green box in the top panel indicates the estimated statistical errors for net-protons, from the RHIC BESII 
program~\cite{besII}. }
\label{fig:net}
\end{SCfigure}

Motivated by these considerations, STAR has studied the kurtosis times the variance ($\kappa$$\sigma^2$) of net-proton (a proxy for net-baryon) and net-charge distributions to search for the critical point~\cite{netp2014,netq2014}. Figure~\ref{fig:net} shows the $\kappa$$\sigma^2$ for mid-rapidity net-proton (top panel)~\cite{netp2014} and net-charge (bottom panel)~\cite{netq2014} distributions in Au+Au collisions as a function of colliding energy for two different collision centralities, 0-5\% (circles) and 70-80\% (squares)\footnote{Note that these protons and anti-protons are measured within the kinematic region: $|y|\le 0.5, 0.4\le p_T \le 0.8$GeV/c.  The transverse moment of protons can be extended up to $p_T \sim$1.6 GeV/c if the time-of-flight information is included in the analysis.}.  In the top panel, the net-proton $\kappa$$\sigma^2$ values for the 0-5\% centrality selection at $\sqrt{s_{\mathrm {NN}}}$ = 19.6 and 27GeV are observed to deviate from: (a) the values from 70-80\% peripheral collisions which are expected to create small systems and do not show significant bulk properties; (b) Poisson and hadron resonance gas expectation values, which would correspond to uncorrelated emission and are close to unity; (c) transport-model-based UrQMD~\cite{urqmd98} calculations in which one does not experience a phase transition. A possible non-monotonic variation of the $\kappa$$\sigma^2$ of the net-proton distribution is not excluded by the existing STAR data. High event statistics for
collisions below 20 GeV in the second phase of the beam energy scan (BESII) will help clarify these issues. The projection of the statistical error from BESII is shown as the hatched band in the top panel.

\subsection{Results on global chiral effects}

\begin{SCfigure}
\includegraphics*[width=8.cm]{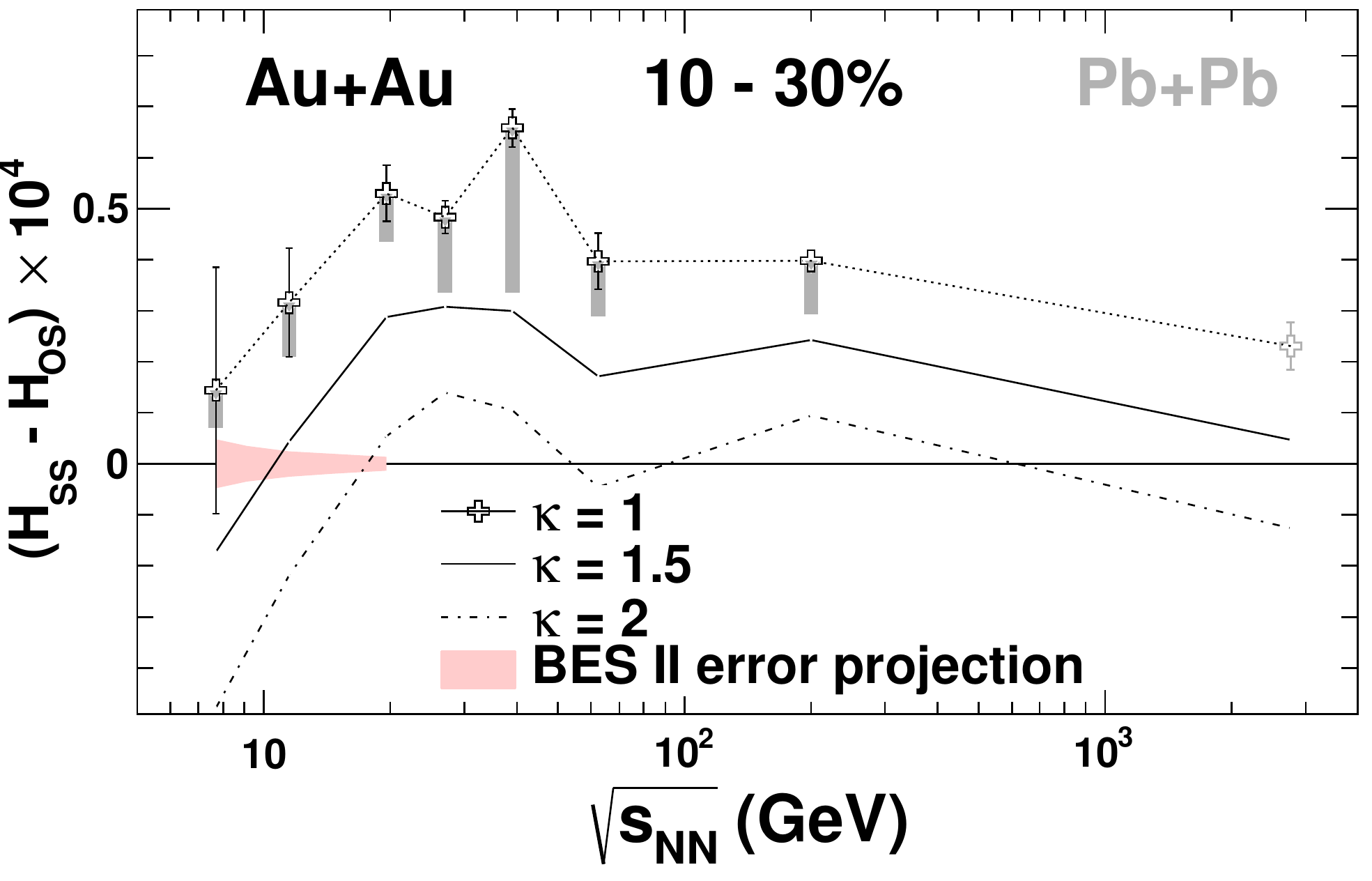}
\caption{Charge separation shown as a function of collision energy for the10-30\% Au+Au collisions. The default values (dotted curves) are from H$^{\kappa =1}$, and the solid (dash-dot) curves are obtained with $\kappa$ = 1.5 ($\kappa$ = 2). For comparison, the results for Pb+Pb collisions at 2.76 TeV are also shown \cite{acme1}.  The vertical asymmetric bands represent the systematic errors and the colored band indicates the statistical errors from the proposed RHIC BESII program. }
\label{fig:cme}
\end{SCfigure}

QCD is a non-Abelian gauge theory for strong interactions and the corresponding global topology is related to the imbalance of chirality. Combined with the external magnetic field, one observes the Chiral Magnetic Effect (CME) where the electric charges are separated along the external magnetic field~\cite{dima2014}. Since the global chiral effect is an intrinsic property of QCD the experimental observation has been challenging. The STAR experiment has employed a three-point correlator method to study this fundamental physics~\cite{svoloshin2004}. Experimentally, charge separation along the external magnetic field, possibly due to the CME effect, has been observed with charged hadron correlations in high-energy nuclear collisions~\cite{scme1,scme2,acme1}. The collision energy dependence of the charge separation, after the background subtraction, is shown in Fig.~\ref{fig:cme}. These results are from the mid-central ($10-30$\%) Au+Au collisions~\cite{qyshou}. The systematic errors are shown as asymmetric bands and the colored band indicates the statistical error bars for the results from the proposed RHIC BESII program. As one can see in the figure, when the energy is lower than 11 GeV the charge separation approaches zero.

On the other hand, one does not expect to see such charge separation when the correlation measurements involve neutral hadrons that contain oppositely charged quarks. At this QM conference, STAR reported its first results on $K_S^0$-h$^{\pm}$ and $\Lambda$-$h^{\pm}$ correlation functions~\cite{fzhao}. Again these results are consistent with the CME expectation.

\begin{SCfigure}
\includegraphics*[width=7.cm]{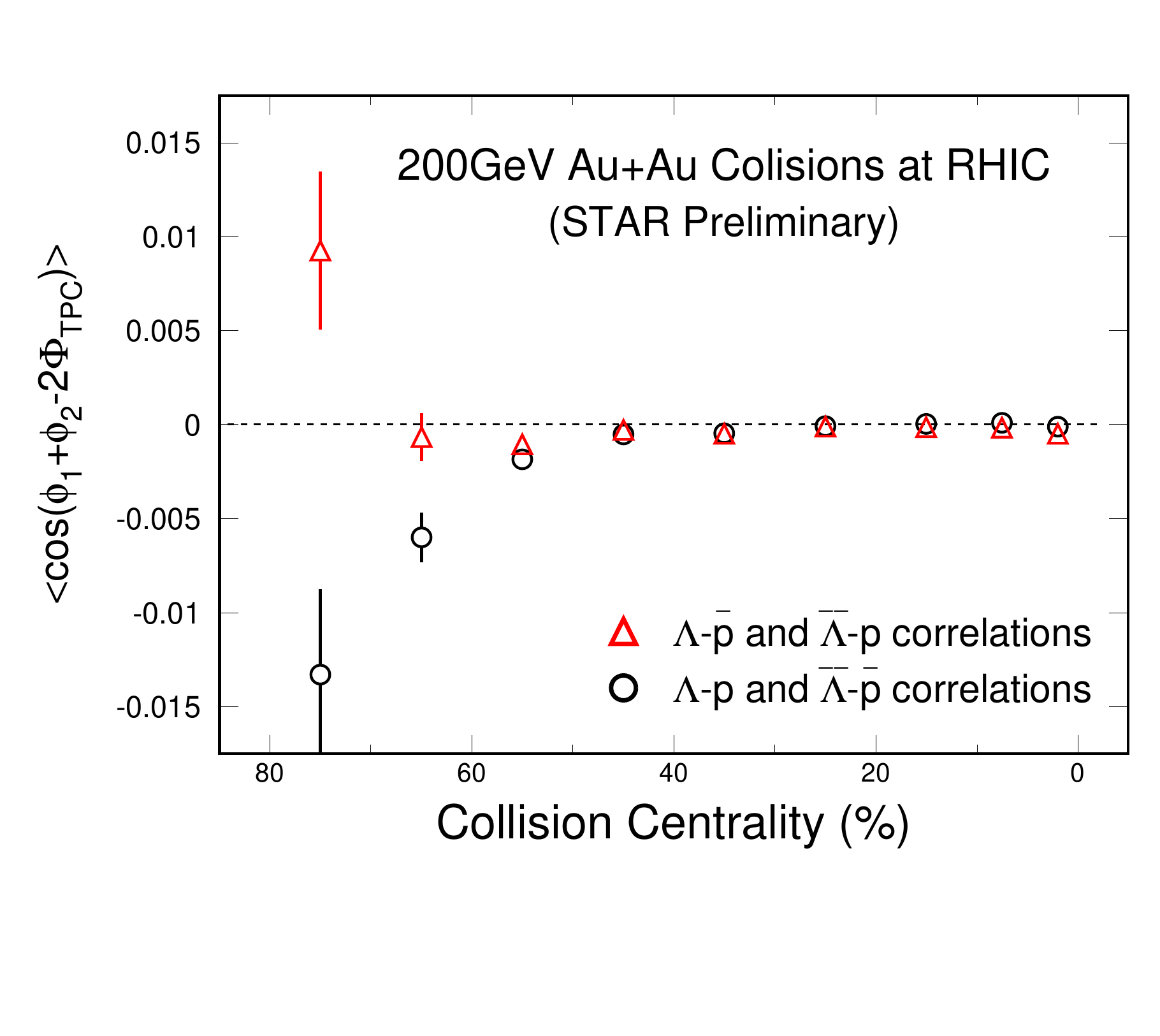}
\vspace{-1.5cm}
\caption{Baryon separation shown as a function of collision centrality from $\sqrt{s_{NN}}=200$GeV Au+Au collisions. Circles and triangles represent the data for the same-baryon-number and opposite-baryon-number correlations, respectively. Error bars are statistical only.}
\label{fig:cve}
\end{SCfigure}

When one replaces the external magnetic field, in the case of the CME,  with the external angular momentum, one expects the so-called Chiral Vortical Effect (CVE). In high-energy nuclear collisions, although the external magnetic field is as strong as $10^{18}$ gauss, it only exists at the very beginning of the collisions and quickly dies away~\cite{dima08}. The angular momentum, on the other hand, is conserved and always stays within the system. If the chiral dynamics does manifest itself in high-energy nuclear collisions, just like the charge separation along the external magnetic field in the CME, one expects the baryon-number separation along the external angular momentum. Figure~\ref{fig:cve} shows the first measurement of the baryon separation, for proton-$\Lambda$ (and their anti-particles) correlation functions with respect to the event plane~\cite{fzhao}. The circles and triangles represent the data for the same-baryon-number (SBN) and opposite-baryon-number (OBN) correlations, respectively. Error bars are statistical only. It is clear from the figure that, as a function of centrality, the SBN behaves differently from that of OBN. Several sources, such as collective flow, resonance decay, or quantum effects other than the CVE may contribute to the observed baryon separation. More detailed studies are called for before one draws further conclusions on this observation. 

\section{Final Remarks on STAR's Near Future Physics Programs}
\label{future}

Although spin physics is not the focus of the conference, STAR's spin program is also very successful in studying the gluon contributions to the proton helicity structure~\cite{spin052014,spin042014}.  Now we turn our attention to the near future heavy-ion physics programs at STAR. Based on the needs of our physics goals, we roughly divide the next few years into three periods:
 
{\bf Period 2014-2016:} The STAR experiment has just successfully commissioned its heavy flavor tracker (HFT) utilizing the state of art MAPS technology~\cite{hqiu} and the Muon Telescope Detector (MTD)~\cite{ljruan09}. These upgrades will enhance STAR's capabilities for heavy flavor measurements including both  open heavy quark hadrons at low transverse momenta and quarkonia. This program focuses on the top energy 200 GeV Au+Au and p+p collisions. Information on the sQGP properties such as the shear viscosity and temperature is expected from these measurements.  

{\bf Period 2018-2020:} RHIC CAD is undertaking an upgrade of the electron cooling. The aim is to increase the collider luminosity by a factor of $3-10$ for the energy range $\sqrt{s_{NN}}=5-20$ GeV, respectively.  In the mean time, the STAR experiment is carrying out two related upgrades~\cite{besII}: (1) The Event Plane Detector (EPD) which will allow an independent determination of the event plane in addition to the TPC, suppressing auto-correlations.  (2) The inner TPC (iTPC) which will greatly enhance the tracking momentum resolution and extend the rapidity range from $|\eta|<1.1$ to $1.7$.  The extended rapidity region is necessary for collectivity measurements at STAR, especially for identified $v_1$ measurement.  In addition, with the moderate forward detector upgrade at STAR~\cite{decadalplan2014}, the precision tracking at high rapidity, $i.e.$ small-x region, will also be necessary for studying initial physics in heavy-ion collisions via the polarized $pp$ and $pA$ collisions. This is also part of the long term future, as discussed in the proposed $eSTAR$ program~\cite{estar}, at RHIC. 

For collisions in the $\sqrt{s_{NN}} \le 20$ GeV region we have observed several interesting changes~\cite{besII} including the measured $\langle p_T \rangle$ for identified hadrons~\cite{lkumar}, the dropping of the $\phi-$meson elliptic flow $v_2$, and the net-proton kurtosis starts to deviate from the monotonic trend, see Fig.~\ref{fig:net}, for example. The proposed RHIC BESII program will focus on this energy region and information of the detailed phase structure will be determined. For the fluctuations analysis, we have reached to fourth order cumulants for net-charge, net-Kaon and net-protons~\cite{asarkar}. In order to pin down the location of the critical point one may need to analyze the data with even higher orders such as the sixth cumulant. In addition, until now we have only considered the Au+Au collisions at RHIC. The chemical freeze-out in small colliding system may be closer to the phase boundary. To avoid the geometry-induced uncertainties in peripheral collisions, one might consider to use central collisions of smaller nuclei, such as Si+Si, Cu+Cu and In+In collisions in the future.   

{\bf Beyond 2020:} As we mentioned above, heavy flavor production at RHIC is important for the study of the sQGP properties. In the top energy $\sqrt{s_{NN}}=200$ GeV Au+Au collisions, however, final-state interactions, $i.e.$ regeneration for $J/\psi$ and recombination for open-charm hadrons, are significant for charm hadron production~\cite{pfzhuang06,rrapp13}.  These facts make the charm hadrons less attractive for studying early dynamics in such collisions. The mass of the bottom quark is around 5 GeV/c$^2$ and it is rarely produced at the RHIC energy. It can serve as a clean probe for studying dynamic evolution in heavy-ion collisions. One could consider to upgrade the HFT MAPS to a much faster ones (HFT') so a trigger can be constructed. It would allow a systematic study of the bottom hadron production at RHIC. This is a unique physics program itself and can be compared to the heavy-ion collision results from the LHC program. In addition, this HFT' heavy flavor physics will be complementary to the proposed sPHENIX~\cite{xche14} jet physics program making the RHIC physics program much stronger and, perhaps, more conceivable.  

{\bf Acknowledgements:} I would like to thank the STAR experiment for the opportunity to present the talk at the QM2014 conference. This work was supported in part by the Office of Science, US Department of Energy under Contract No. DE-AC03-76SF00098 and the National Natural Science Foundation of China under grant No.11221504. 







\begin{multicols}{2}
\bibliographystyle{elsarticle-num}

\end{multicols}

\end{document}